\documentclass[pre,aps,amsmath,amssymb,reprint,floatfix,superscriptaddress,longbibliography]{revtex4-1}

\pdfoutput=1
\usepackage{xcolor}
\usepackage{siunitx}
\usepackage{graphicx}
\usepackage[colorlinks=true,citecolor=blue,urlcolor=blue]{hyperref}
\usepackage{etoolbox} % Add this line to include etoolbox package
\usepackage{bm} % Added for bold math symbols
\usepackage[normalem]{ulem}
\newcommand{\bl}[1]{\textcolor{blue}{#1}}
\newcommand{\new}[1]{\textcolor{black}{#1}}

\let\cedilla\c

\newcommand{\thin}{\mskip0.75mu}

\def\cal#1{\mathcal{#1}}
\def\eqq#1{Eq.~(\ref{#1})}

\def\eq#1{(\ref{#1})}

\def\f#1{Fig.~\ref{#1}}

\def\s#1{Section~\ref{#1}}

\def\c#1{~\cite{#1}}

\def\av#1{\langle #1 \rangle}

\def\beq{\begin{equation}}
\def\eeq{\end{equation}}
\def\bea{\begin{eqnarray}}
\def\eea{\end{eqnarray}}

\def\x{{\bm x}}
\def\tt{{\bm \theta}}

\def\kB{k_{\rm B}}
\def\tf{t_{\rm f}}
\def\kt{\kB T}

\begin{document}

\title{Evolutionary design of thermodynamic logic gates and their heat emission}

\author{Stephen Whitelam}
\email{swhitelam@lbl.gov}

\affiliation{Molecular Foundry, Lawrence Berkeley National Laboratory, 1 Cyclotron Road, Berkeley, CA 94720, USA}

\begin{abstract}
Landauer's principle bounds the heat generated by logical operations, but in practice the thermodynamic cost of computation is dominated by the control systems that implement logic. \new{Complementary metal oxide semiconductor} (CMOS) gates dissipate energy far above the Landauer bound, while laboratory demonstrations of near-Landauer erasure rely on external measurement or feedback systems whose energy costs exceed that of the logic operation by many orders of magnitude. Here we use simulations to show that a genetic algorithm can program a thermodynamic computer to implement logic operations in which the total heat emitted by the control system is of a similar order of magnitude to that of the information-bearing degrees of freedom. Moreover, the computer can be programmed so that heat is drawn away from the information-bearing degrees of freedom and dissipated within the control unit, suggesting the possibility of computing architectures in which heat management is an integral part of the program design.
\end{abstract}
 
\maketitle

\section{Introduction}

Landauer's principle states that the heat emitted upon erasing a one-bit memory can be as little as $\kt \ln 2$, where $T$ is the temperature of the thermal bath\c{landauer1961irreversibility}. However, real computing hardware operates at energy scales far above this bound. CMOS hardware implements simple logic operations with dissipation of order $10^4$ to $10^6 \, \kt$\c{horowitz20141,datta2022toward}. Laboratory realizations of information engines and feedback-controlled particles can approach the Landauer bound if we consider only the degree of freedom that stores the bit\c{berut2012experimental,jun2014high,hong2016experimental,dago2021information,chattopadhyay2025landauer}, but require cameras, amplifiers, digitizers, or feedback-control processors whose heat emission exceeds that of the information-bearing degrees of freedom by many orders of magnitude. For instance, floating-point arithmetic used in digital control devices requires upwards of $10^8 \, \kt$ {\em per operation}\c{horowitz20141}, and many such operations are required to implement a control protocol. The thermodynamic cost of a logical operation is currently dominated by the control apparatus, not by the information register.

Here we use simulations to show that it is possible to program a thermodynamic logic device whose control apparatus dissipates heat of a similar order of magnitude to the information register it controls. Thermodynamic computing is an emerging field that aims to do computation using noisy physical devices that evolve in time according to the Langevin equation\c{conte2019thermodynamic,hylton2020thermodynamic, hylton2020thermodynamicnn,wimsatt2021harnessing,boyd2022shortcuts,aifer2024thermodynamic,melanson2025thermodynamic, whitelam2026generative}. We consider a thermodynamic device that consists of information-storing units coupled to computational units. We use a genetic algorithm to train the device to perform erasure and XOR operations under different objectives: maximizing fidelity alone; minimizing total heat at fixed fidelity; and minimizing heat associated with the information register at fixed fidelity. We show that the controller can act as an internal feedback mechanism, similar to a Maxwell demon, that reduces or even reverses heat flow from the information-storing degrees of freedom.

The total heat emission of a computational device is constrained by its speed of operation and fidelity by fundamental results of statistical mechanics and stochastic thermodynamics\c{bennett1982thermodynamics,seifert2012stochastic,proesmans2020finite,freitas2021stochastic,konopik2023fundamental,yoshino2023thermodynamics,wolpert2024stochastic,klinger2025minimally,helms2025stochastic,rolandi2026energy, tang2026momentum}. \new{In general, overdamped Langevin systems exhibit a tradeoff between program run time, computational progress, and heat dissipation\c{dechant2019thermodynamic}. Here we fix the run time by fixing the observation time of the device, and explore the tradeoff between computational progress, measured through the fidelity of the logic operation, and the heat dissipated.} Our results indicate that there is considerable freedom to arrange {\em where} in a device heat is produced. This freedom allows heat to be moved away from information-bearing elements and into control elements, suggesting computing architectures in which heat management is an integral part of the program design.

\new{Note that logic devices have also been developed in the field of {\em thermal} computing, in which heat currents or temperature differences are used directly as computational signals, often within deterministic device architectures \c{wang2007thermal,hamed2020nanothermomechanical,xu2024electric}. Here, by contrast, we consider devices that operate using the stochastic, noise-driven dynamics characteristic of thermodynamic computing.}

In \s{model} we introduce a computational model of a thermodynamic logic device consisting of information-bearing thermodynamic degrees of freedom coupled to computational thermodynamic degrees of freedom. In \s{training} we describe how to adjust the couplings between these units so that the device implements erasure or XOR gates. The gate inputs are the initial logical states of the visible units; the gate output is the final-time logical state of the first visible unit, which results from the natural (Langevin) dynamics of the coupled thermodynamic system. In this mode of operation we use the thermodynamic computer as a {\em Langevin computer}, designed to run a specified program using finite-time Langevin trajectories. In \s{results} we assess the performance of the trained device, and show that it is possible to program the thermodynamic computer to implement the specified logic with different characteristics of heat emission. We conclude in \s{conclusions}.

\section{Model of a thermodynamic logic device} 
\label{model}

Our model of a thermodynamic logic device, which comprises information-storing thermodynamic degrees of freedom coupled to a thermodynamic computer, consists of $N=N_{\rm v}+N_{\rm h}$ classical, real-valued fluctuating degrees of freedom $\x=\{x_i\}$. The system has $N_{\rm v}$ {\em visible} degrees of freedom, the information-storing elements, and $N_{\rm h}$ {\em hidden} degrees of freedom, the computational elements. The total potential energy of the system is
\bea
\label{pot}
V_\tt(\x) &=& \sum_{i \in {\rm v}}  \left(J_2^{(\rm v)} x_i^2+J_4^{(\rm v)} x_i^4\right)+ \sum_{(ij)} J_{ij} x_i x_j \\
 &+&\sum_{i \in {\rm h}}  \left(J_2^{(\rm h)} x_i^2+J_4^{(\rm h)} x_i^4\right)
+\sum_{i \in {\rm h}} b_i x_i, \nonumber
\eea
where $\tt = \{\{J_{ij}\},\{b_i\}\}$ are the adjustable parameters. 

The first sum in \eq{pot} runs over all visible degrees of freedom. The parameters $J_2^{(\rm v)}=-7\, \kt$ and $J_4^{(\rm v)}=\kt$ impose an on-site potential with a double-well form, shown schematically in blue in \f{fig1}(a) (the on-site potential has minima at $x_i=\pm x_0$, where $x_0^2= -J_2^{(\rm v)}/(2 J_4^{(\rm v)})=3.5$, and a barrier height of $(J_2^{(\rm v)})^2/(4 J_4^{(\rm v)})=12.25 \, \kt$). These degrees of freedom are similar to the {\em p-spins} of probabilistic computers, which are often realized in hardware by magnetic memories\c{kaiser2021probabilistic,aadit2022massively,misra2023probabilistic}. We associate logical states $S_i(t) = ({\rm sign}(x_i(t))+1)/2=0$ and 1 with positive and negative values $x_i(t)$ of the visible units, respectively. 
\begin{figure} 
   \centering
   \includegraphics[width=\linewidth]{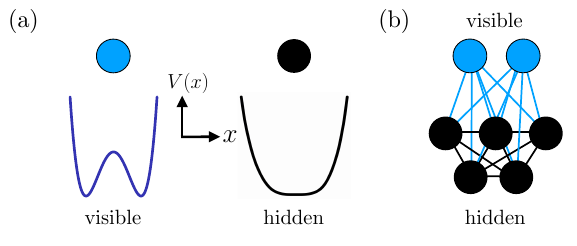} 
   \caption{(a) In this paper we consider simulations of a thermodynamic logic device built from {\em visible} and {\em hidden} units. Visible units feel a double-well potential described by the first sum in \eqq{pot}. These are information-storing units: logical states 0 and 1 correspond to the left- and right-hand portions of the double-well potential. Hidden units feel a (non-quadratic) single-well potential described by the third sum in \eqq{pot}; these are computational units. (b) We consider $N_{\rm v}$ visible units (blue) and $N_{\rm h}$ hidden units (black). There are connections $J_{ij}$ between visible and hidden units (blue), and between hidden units (black), described by the second sum in \eqq{pot}. These connections, and the hidden-unit biases $b_i$ described by the fourth sum in \eqq{pot}, are adjusted by genetic algorithm so that the logic device performs erasure (reset-to-zero) or XOR operations. For erasure and XOR we take $N_{\rm v}=1$ and 2, respectively, and in both cases we take $N_{\rm h}=10$. }
   \label{fig1}
\end{figure}

The second sum in \eq{pot} runs over all distinct pairwise interactions. We consider connections between all visible units and all hidden units (shown blue in \f{fig1}(b)), and between all hidden units (shown in black in \f{fig1}(b)). There are no connections between visible units. The couplings $J_{ij}$ are part of the set of adjustable parameters $\tt$. The bilinear interaction $x_i x_j$ is a simple choice modeled on the interactions used in existing thermodynamic computers\c{aifer2024thermodynamic,melanson2025thermodynamic}.

The third sum in \eq{pot} runs over all hidden degrees of freedom, which are the system's computational elements. The parameters $J_2^{(\rm h)}= J_4^{(\rm h)}=\kt$ impose an on-site potential with a single-well form, shown schematically in black in \f{fig1}(a). These degrees of freedom are similar to the {\em s-units} of thermodynamic computers, which have been realized in hardware by RLC circuits\c{aifer2024thermodynamic,melanson2025thermodynamic}. We consider a non-quadratic potential with $J_4^{(\rm h)} \neq 0$: the resulting network of hidden units has the ability to express a nonlinear function of its inputs, with similar expressive power to a digital neural network\c{whitelam2026nonlinear}.

The final sum in \eq{pot} runs over all hidden units. The bias parameters $b_i$ are part of the set of adjustable parameters $\tt$.

The potential energy \eq{pot} describes a thermodynamic system with $N_{\rm v}$ visible or information-storing degrees of freedom, shown in blue in \f{fig1}(b), and $N_{\rm h}$ hidden or computational degrees of freedom, shown in black in \f{fig1}(b). These degrees of freedom are coupled through the interaction terms $J_{ij}$. Our aim is to adjust the couplings $J_{ij}$ and the hidden-unit biases $b_i$ so that the visible degrees of freedom carry out a particular program.

\new{In order for the computer to express a nonlinear mapping between input and output it is important that the response of each thermodynamic degree of freedom to external stimulus be nonlinear, which we accomplish using the quartic terms in \eq{pot}\c{whitelam2026nonlinear}. The quartic form is the simplest nonlinear choice. More generally, we expect the qualitative behavior reported here to persist for a broad class of nonlinear potentials, although the detailed dynamics and heat flows would depend on the particular form chosen. In this sense, changing the nonlinear potential is analogous to changing the activation function in a neural network: the details of the computation change, but the system remains generically expressive.}

We assume that the system is in contact with a thermal bath, and evolves in time according to the overdamped Langevin dynamics
\beq
\label{lang1}
\dot{x}_i= -\mu \thin \partial_i V_\tt(\x)  + \sqrt{2 \mu \thin \kt} \, \eta_i(t).
\eeq 
 The first term on the right-hand side of \eqq{lang1} is the force arising from the computer's potential energy $V_\tt(\x)$, where $\partial_i \equiv \partial/\partial x_i$. The second term models thermal fluctuations: $\kt$ is the thermal energy scale, and the Gaussian white noise terms satisfy $\av{\eta_i(t)}=0$ and $\av{\eta_i(t) \eta_j(t')} = \delta_{ij} \delta(t-t')$. The mobility parameter $\mu$ sets the basic time constant of the computer. For the thermodynamic computers of Refs.\c{aifer2024thermodynamic,melanson2025thermodynamic}, $\mu^{-1}$ is of order a microsecond. For damped oscillators made from mechanical elements\c{dago2021information} or Josephson junctions\c{ray2023gigahertz,pratt2025controlled}, $\mu^{-1}$ is of order a millisecond or a nanosecond, respectively.

A single dynamical trajectory of the system is constructed as follows. We take the initial states $x_i(0)$ of the visible units to correspond to the thermal equilibrium associated with the isolated visible-unit potential, the double-well form shown schematically in blue in \f{fig1}(a). The initial logical states $S_i(0)$ of the visible spins can therefore take values 0 or 1. We take the initial states $x_i(0)$ of the hidden units to be zero. We allow the system to evolve for time $\tf=1$ (in units of $\mu^{-1}$), and we record the values $x_i(\tf)$ and logical states $S_i(\tf)=0,1$ of the visible units at time $\tf$. 

\new{Thus, for a given realization of the couplings and biases, the logic operation proceeds autonomously. The initial logical states of the visible units define the gate input. These visible units interact with the hidden units through the potential energy \eq{pot}, and evolve dynamically according to the Langevin equation \eq{lang1}. The output of the device is the final-time logical state of the first visible unit.}

\section{Training the thermodynamic computer to implement logic gates}
\label{training}

Our aim is to train the thermodynamic computer to implement two types of logic gate, erasure (reset-to-zero), and XOR~\footnote{We also verified that the genetic algorithm can train the logic device to implement NAND with unit fidelity. NAND is significant because it is computationally universal, and can be used to construct any Boolean computation.}. To study erasure we consider $N_{\rm v}=1$ visible unit and $N_{\rm h}=10$ hidden units. To study XOR we consider $N_{\rm v}=2$ visible units and $N_{\rm h}=10$ hidden units. We take the inputs to each gate to be the initial logical states of the visible spins, and the output of each gate to be the final logical state of the first visible spin (which is the only visible spin in the case of erasure). 

For erasure we want the final logical state $S_1(\tf)$ of the visible spin to be 0 regardless of its initial logical state. For XOR we want the final state $S_1(\tf)$ of the first visible spin to be 1 if $S_1(0) \neq S_2(0)$, and 0 if $S_1(0) = S_2(0)$. XOR is a more difficult computational task than erasure, because it is not linearly separable\c{minsky1969perceptrons}.

We also wish to produce versions of these gates that balance program fidelity and heat emission. We measure the heat emitted over the course of a trajectory as 
\beq
Q=V_\tt(\x(\tf))-V_\tt(\x(0)),
\eeq
with negative values indicating heat emitted to the thermal bath. We also consider the heat emitted by the visible degrees of freedom,
\beq
Q_{\rm v}=V_\tt^{(\rm v)}(\x(\tf))-V_\tt^{(\rm v)}(\x(0)),
\eeq
where
\beq
V_\tt^{(\rm v)}(\x) = \sum_{i \in {\rm v}}  \left(J_2^{(\rm v)} x_i^2+J_4^{(\rm v)} x_i^4\right)+ \sum_{i \in {\rm v}, j \in {\rm h}} J_{ij} x_i x_j 
\eeq
is the potential energy associated with visible degrees of freedom and visible-hidden couplings, both of which are shown blue in \f{fig1}(b). The visible-unit heat $Q_{\rm v}$ is the register heat considered in studies of low-energy logic gates; the total heat, which includes the heat associated with the control elements of the system, is not usually considered in those studies.

We train the logic device using a mutation-only genetic algorithm\c{holland1992genetic,mitchell1998introduction,such2017deep,whitelam2023demon} designed to minimize an order parameter $\phi$. The order parameter (described below) is calculated using $10^4$ independent dynamical trajectories of the thermodynamic logic device. Starting from an ensemble of 50 devices, each with randomly-chosen parameters $\theta_i \sim {\cal N}(0,\sigma^2)$, where $\sigma = 0.01$, we evaluate $\phi$ for each device. We retain the devices with the 5 smallest values of $\phi$, and create a new ensemble of 50 by drawing with replacement from this set of 5 and perturbing each parameter of each device by a Gaussian random number, $\theta_i \to \theta_i + \epsilon_i$ , where $\epsilon_i \sim {\cal N}(0,\sigma^2)$. This procedure is repeated until a desired performance has been achieved. 
\begin{figure} 
   \centering
   \includegraphics[width=\linewidth]{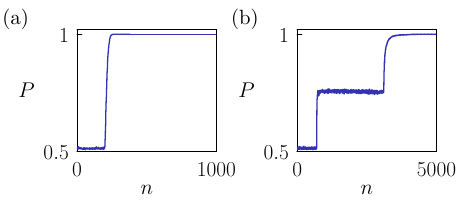} 
   \caption{Program fidelity $P$ as a function of evolutionary time $n$ for (a) erasure and (b) XOR. Here the genetic algorithm is instructed to minimize the order parameter $\phi_1$, \eqq{phi1}.}
   \label{fig2}
\end{figure}

We consider three evolutionary order parameters $\phi$. The first, $\phi_1$, is designed to maximize the fidelity of the program (erasure or XOR). Let $P$ be the probability of success of the protocol, measured over $10^4$ trajectories. Then
\begin{align}
\phi_1 &=
\begin{cases}
\Delta+\Lambda_0, & P<P_0,\\
1-P, & P \ge P_0.
\end{cases}
\label{phi1}
\end{align}

Here $P_0=0.9$ is a threshold fidelity. The quantity $\Delta = \av{|x_1(\tf)-x_{\rm target}|}$, where the angle brackets denote the mean over $10^4$ independent trajectories, and $x_{\rm target} = (2 S_{\rm target}-1) |x_0|$, where $S_{\rm target}=0,1$ is the desired logical state of the first visible unit at the end of the trajectory (recall that $\pm x_0$ are the positions of the visible-spin potential minima). The ultimate goal of \eqq{phi1} is to minimize $1-P$, and hence maximize the protocol fidelity $P$. The first clause in \eq{phi1} is needed because, initially, $P$ does not change significantly as the parameters of the logic device are changed. The first clause encourages the first visible unit to end the trajectory increasingly near to the center of the well corresponding to the target logical state, which eventually leads to an increase of $P$. Once $P$ exceeds 0.9, the second clause becomes active and encourages maximization of $P$. The constant $\Lambda_0=500$ ensures that the value of the first clause is always larger than that of the second, so ensuring that the second clause represents the ultimate goal of the program.

The second order parameter, $\phi_2$, is chosen to balance protocol fidelity and total heat emission, and is
\begin{align}
\phi_2 &=
\begin{cases}
\Delta+\Lambda_0, & P<P_0,\\
1-P+\Lambda_1, & P_0 \le P < P_1,\\
-\,c\,\av{Q}, & P \ge P_1.
\end{cases}
\label{phi2}
\end{align}
Here $\Lambda_0=500, \Lambda_1=100$, $P_0=0.9$, $P_1 = 1-10^{-3}$, and $c=10^{-3}$. The first two clauses of \eq{phi2} are similar to those of $\phi_1$, encouraging maximization of protocol fidelity up to a fidelity of 99.9\%. Upon reaching this threshold, the final clause becomes active and encourages minimization of the total emitted heat. Thus minimizing $\phi_2$ minimizes heat emission in a logic device that has achieved at least 99.9\% fidelity (the constants $\Lambda_0$, $\Lambda_1$, and $c$ enforce a clear separation of scales between the three clauses, but their precise numerical values are otherwise unimportant).
\begin{figure*} 
   \centering
   \includegraphics[width=\linewidth]{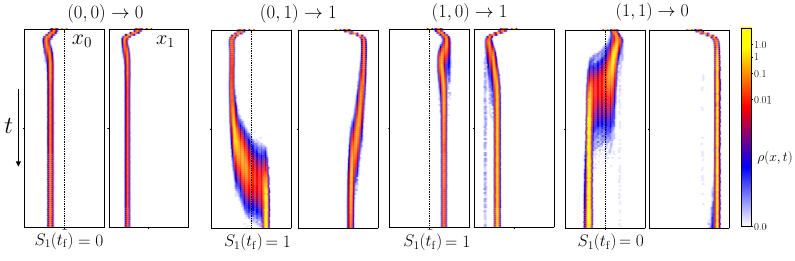} 
   \caption{State-time probability densities $\rho(x_1,t)$ and $\rho(x_2,t)$ for the visible spins under the XOR program produced by minimizing $\phi_1$. The left (right) half of each box corresponds to logical state 0 (1); for the first visible spin, the dividing line is shown dotted. The inputs to the program are the logical states $S_1(0),S_2(0)$ of the two visible units at the initial time (top). The output of the program is the logical state $S_1(\tf)$ of the first visible unit at the final time (bottom). The four panels show the four input cases: (0,0) and (1,1) produce an output of 0, while (0,1) and (1,0) produce an output of 1. Probability densities are measured over $10^4$ independent trajectories. }
   \label{fig3}
\end{figure*}

The third order parameter, $\phi_3$, is chosen to balance protocol fidelity and visible-unit heat emission, and is
\begin{align}
\phi_3 &=
\begin{cases}
\Delta+\Lambda_0, & P<P_0,\\
1-P+\Lambda_1, & P_0 \le P < P_1,\\
-\,c\,\av{Q_{\rm v}}, & P \ge P_1.
\end{cases}
\label{phi3}
\end{align}
Here $\Lambda_0=500, \Lambda_1=100$, $P_0=0.9$, $P_1 = 1-10^{-3}$, and $c=10^{-3}$. Minimizing the order parameter $\phi_3$ minimizes the heat emission associated with the visible degrees of freedom in a logic device that has achieved at least 99.9\% fidelity.

\new{Once the genetic algorithm has determined the device parameters, these could in principle be implemented within a physical realization of a thermodynamic computer. In such a setting, the programmable couplings and biases defining the computer's potential energy would correspond to tunable interactions (e.g. resistances or capacitances) as used in existing thermodynamic computers\c{aifer2024thermodynamic,melanson2025thermodynamic}.}

\section{Results}
\label{results}

 \f{fig2} shows the program fidelity $P$ as a function of evolutionary time $n$ for (a) erasure and (b) XOR programs. Here the genetic algorithm is used to minimize $\phi_1$, which maximizes program fidelity. Both programs reach a fidelity of 1, confirming that it is possible to train a thermodynamic computer to perform logic operations with high fidelity. XOR takes longer to train, and plateaus at a fidelity of $3/4$ for several generations, showing that the genetic algorithm takes time to identify a program that can cope with all input cases.
 
 \begin{table}[h]
\centering
\label{tab1}
\begin{tabular}{lccc}
\hline
process, o.p. & $1-P$ & $-\av{Q}$ & $-\av{Q_{\rm v}}$ \\
\hline
erasure, $\phi_1$
& \bl{0} & $846\, (15.9)$ & $157\, (28.1)$ \\

erasure, $\phi_2$
& $\bl{1.7 \times 10^{-3}}$ & $\bl{77 \, (3.5)}$ & $44\, (4.3)$ \\

erasure, $\phi_3$
& $\bl{0.9 \times 10^{-3}}$ & $2387 \, (6.45)$ & $\bl{-8 \, (2.96)}$ \\

XOR, $\phi_1$
& $\bl{1.3 \times 10^{-4}}$ & $3407 \, (334)$ & $1130 \, (1104)$ \\

XOR, $\phi_2$
& $\bl{1.4 \times 10^{-3}}$ & $\bl{1270 \, (257)}$ & $997 \, (442)$ \\

XOR, $\phi_3$
& $\bl{1 \times 10^{-3}}$ & 3366  (80.0) & \bl{134  (62.1)} \\
\hline
\end{tabular}
\caption{Error probability $1-P$, mean heat $\av{Q}$, and mean visible-unit heat $\av{Q_{\rm v}}$ for erasure and XOR programs trained by minimizing the three order parameters $\phi_1$, $\phi_2$, and $\phi_3$. Heat is measured in units of $\kt$. Minimizing $\phi_1$ maximizes $P$, while minimizing $\phi_2$ and $\phi_3$ minimizes mean heat emission and mean visible-unit heat emission, respectively, for a program with a fidelity of $99.9\%$. The quantities that contribute to each order parameter are shown in \bl{blue}. Programs were trained using $10^4$ trajectories per order-parameter evaluation, and the table was produced by evaluating each of the trained programs using $10^6$ trajectories. \new{Standard deviations of $\av{Q}$ and $\av{Q_{\rm v}}$ are shown in parentheses.}}
\end{table}

Note that fidelity $P$ is measured during training using $10^4$ independent trajectories, meaning that we can expect the true program error rate to be about 1 part in $10^4$. In Table I we assess the fidelity of the erasure and XOR programs produced by minimizing $\phi_1$ using $10^6$ independent trajectories. The error rate for XOR is indeed about 1 in $10^4$, while no errors are detected for erasure.

\begin{figure*} 
   \centering
   \includegraphics[width=0.9\linewidth]{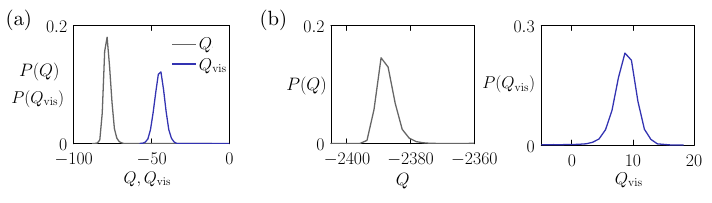} 
   \caption{Histograms of total heat (black) and visible-unit heat (blue) emitted by the erasure program trained using the order parameters (a) $\phi_2$ and (b) $\phi_3$. Upon moving from one order parameter to the other, the total heat emission increases markedly, and the visible units go from emitting to absorbing heat. Histograms were taken over $10^4$ independent trajectories.}
   \label{fig4}
\end{figure*}

\new{The logic device operates through the autonomous dynamics specified by \eqq{lang1}, using the programmed couplings of \eqq{pot}, rather than through a hand-designed sequence of operations. To visualize the dynamics of the computation in the subspace of the information-bearing degrees of freedom, we present Fig. 3. There we show show state-time probability densities $\rho(x_1,t)$ and $\rho(x_2,t)$ for the XOR program} produced by minimizing $\phi_1$. The initial logical states of the two visible units set the input to the program, while the final logical state of the first visible unit represents the output of the program. Note that the color scale is logarithmic, with yellow and blue representing typical and rare events, respectively.  

In Table I we also show the mean total heat emission and mean visible-unit heat emission for erasure and XOR programs. Under the order parameter $\phi_1$, no consideration is paid to heat. As a result, the heat emission in both cases is considerable, more than 800 $\kt$ for erasure, and over four times that for XOR. 

However, under the order parameter $\phi_2$, the genetic algorithm identifies programs that minimize total heat, provided that the program fidelity reaches 99.9\%. Table I shows that the consequent small reduction in program fidelity is accompanied by considerable heat savings. When programmed by $\phi_2$ as opposed to $\phi_1$, the total heat of erasure drops by more than a factor of 10, to $77\, \kt$. The corresponding reduction for XOR is a factor of about 2.7, a savings of more than $2000 \, \kt$.

Significantly, the genetic algorithm can also identify programs that reduce heat emission from selected parts of the device. Under the order parameter $\phi_3$, the genetic algorithm identifies programs that minimize visible-unit heat, provided that the program fidelity exceeds 99.9\%. Table 1 shows that under XOR, when programmed by $\phi_3$ as opposed to $\phi_2$, the visible units emit more than seven times less heat (134 $\kt$ compared to 997 $\kt$). To compensate, more than twice as much total heat is emitted. 

For erasure, the difference is even more significant. When programmed by $\phi_3$ as opposed to $\phi_2$, the total heat increases markedly, from $77\, \kt$ to $2387\, \kt$. However, the visible units change from emitting $44\, \kt$ of heat to {\em absorbing} $8\, \kt$.

\new{For both erasure and XOR, minimizing visible heat emission, as prescribed by the order parameter $\phi_3$, incurs an energetic cost that results in greater total heat emission than under $\phi_2$, which encourages the system to minimize total heat emission. The total heat emitted under $\phi_1$ may be either greater or less than that emitted under $\phi_3$, because $\phi_1$ encourages the system to maximize fidelity only.}

Our previous work showed that a neural network, a digital Maxwell demon, could be trained by genetic algorithm to extract energy from a fluctuating system if supplied with measurements of that system\c{whitelam2023demon,whitelam2023train}. Here the digital demon is replaced by the hidden units of the thermodynamic device. These units do not make measurements explicitly, but instead detect the state of the visible units through their mutual energetic couplings. The thermodynamic control system is much more efficient energetically than a neural network: the total heat cost of $2387\, \kt$ is small compared to even one neural-network multiply-accumulate operation, which on conventional digital hardware exceeds $10^8\, \kt$\c{horowitz20141}.

In \f{fig4} we show histograms of heat and visible-unit heat emitted for the erasure program trained using the order parameter (a) $\phi_2$ and (b) $\phi_3$. Upon moving from one order parameter to the other, the total heat emission increases markedly, while the visible units go from emitting heat to absorbing it.

\section{Conclusions}
\label{conclusions}

We have shown that a thermodynamic logic device can be programmed by genetic algorithm so that the heat emitted by its control degrees of freedom is of the same order of magnitude as that of its information-bearing degrees of freedom, in contrast to the operation of near-Landauer logic experiments controlled by digital devices. When trained for fidelity alone, the thermodynamic device implements erasure and XOR operations with high fidelity and with substantial heat dissipation. When trained instead to minimize total heat at fixed fidelity, it can perform the same logical tasks at slightly reduced fidelity but with markedly reduced heat emission. Most significantly, when the training objective penalizes visible-unit heat at fixed fidelity, the controller degrees of freedom can substantially reduce or even reverse heat flow from the information-storing degrees of freedom. For the erasure operation the visible unit can on average {\em absorb} heat, while the hidden units dissipate it. 

\new{In the present paper we focus on the thermodynamic cost of operating an already-programmed device, but it is instructive to roughly estimate the cost of training the thermodynamic device on a digital computer. Evaluating forces using the potential \eq{pot} requires of order $10^2$ multiply-accumulate (MAC)-like operations per time step, and there are of order $10^3$ time steps per trajectory. We used $50 \times 10^4$ trajectories per generation, and trained the XOR program for 5000 generations. Training therefore incurs a computational cost of about $10^{14}$ MACs. On standard hardware one MAC costs about $1 \, {\rm pJ} \approx 2 \times 10^8 \, \kt$\c{horowitz20141}, giving an energy cost of about $10^{22}\, \kt$, or about $100 \, {\rm J}$. However, the device is trained once and may then be used repeatedly, and so the training cost can be amortized over many logic operations. Moreover, if we wish to account for all training costs then we must also include the cost of fabricating the device, which would substantially exceed the cost of training it.}

\new{The operation of Langevin devices is governed by the constraints of stochastic thermodynamics. Our results show that, within these constraints, it is possible to control {\em where} in a device heat is produced, suggesting computing architectures in which heat management is part of the program specification. In modern computer hardware, localized hot spots can cause logic errors and reduced device lifetime\c{datta2022toward}. Heat control could therefore be valuable when information-bearing elements are especially sensitive to local heating or temperature fluctuations: heat could be moved toward parts of the device engineered for heat dissipation. Our results suggest that the spatial distribution of heat production within a thermodynamic computer can itself be programmed, providing a potentially useful design principle for future thermodynamic or stochastic computing architectures.}

\section{Acknowledgments} I thank Neil Kakhandiki for discussions. This work was done at the Molecular Foundry, supported by the Office of Science, Office of Basic Energy Sciences, of the U.S. Department of Energy under Contract No. DE-AC02-05CH11231, and partly supported by US DOE Office of Science Scientific User Facilities AI/ML project ``A digital twin for spatiotemporally resolved experiments''.

%\bibliography{short_bib}


\end{document}